\documentclass{article}
\usepackage{amsmath,amsthm}
\usepackage{amssymb,latexsym}
\usepackage[mathscr]{eucal}
\usepackage{setspace}
\usepackage{graphics}
\usepackage{array}

\newcommand{\pa}{\partial}

\newcommand{\pr}{\prime}
\newcommand{\dpr}{\prime\prime}
\newcommand{\al}{\alpha}

\newcommand{\lp}{\left(}
\newcommand{\rp}{\right)}
\newcommand{\lb}{\left[}
\newcommand{\rb}{\right]}

\newcommand{\rc}{\right\}}

\newcommand{\beq}{\begin{equation}}
\newcommand{\eq}{\end{equation}}

\newcommand{\fhalf}{\frac{1}{2}}

\newcommand{\nhat}{\bf\hat{n}}
\newcommand{\that}{\bf\hat{t}}

\newcommand{\bhat}{\bf\hat{b}}
\newcommand{\tex}{\textit}

\begin{document}

\title{\textbf{Effect of Slipping Motion on the Hasimoto Soliton on a Vortex Filament in Self-Induced Motion: An Exact Solution}}         
\author{B. K. Shivamoggi\\
University of Central Florida\\
Orlando, FL 32816-1364, USA\\
}        
\date{}          
\maketitle

\noindent \large{\bf Abstract} \\ \\

A vortex filament immersed in a non-ideal fluid, thanks to viscous diffusion, experiences a slipping motion with respect to the fluid. In recognition of this, in this paper, the effect of this slipping motion on the Hasimoto soliton propagating on a vortex filament is investigated, and an \tex{exact} solution is given to describe this process. A strong slipping motion is shown to prevent the existence of the Hasimoto soliton. The critical slipping speed (above which the Hasimoto soliton fails to exist) is shown to be equal to the torsion.

\pagebreak

\noindent\large\textbf{1. Introduction}\\

Da Rios \cite{dar} did an asymptotic calculation, called the \tex{ local induction approximation} (LIA), to resolve the singularity due to the neglect of the finite vortex core size and give the leading-order behavior of the self-induced flow velocity formula for a vortex filament. This formulation led to a set of two coupled equations (\tex{Da Rios equations}) governing the inextensional motion of a vortex filament in an ideal fluid in terms of time evolution of it intrinsic geometric parameters - curvature and torsion. The LIA was reinvented by Arms and Hama \cite{arm} and the Da Rios equations were rediscovered by Betchov \cite{bet} decades later. Hasimoto \cite{has} gave an ingenius development to combine the \tex{Da Rios-Betchov equations} elegantly to give a nonlinear Schrodinger equation. The single-solition solution of this equation (Zakharov and Shabat \cite{zak}) describes an isolated loop of helical twisting motion along the vortex line\footnote{It may be mentioned that computation of the inverse Hasimoto transformation mapping of the intrinsic geometric parameters space onto the extrinsic vortex filament coordinate space is a highly non-trivial task (Sym \cite{sym}, Aref and Flinchem \cite{are}). A reformulation of the Da Rios-Betchov equations in the \tex{extrinsic} vortex filament coordinate space was shown (Shivamoggi and van Heijst \cite{shi}) to provide a useful alternative approach in this regard.} (Majda and Bertozzi \cite{maj}).

A vortex filament immersed in a non-ideal fluid (where viscous diffusion materializes) experiences a slipping motion with respect to the fluid. On the other hand, in a superfluid, vortex lines, which are frozen in the superfluid component, slip past the normal fluid (Donnelly \cite{don}). The purpose of this paper is to investigate the effect of this slipping motion on the Hasimoto soliton; an \tex{exact} solution is given to describe this process.

\vspace{.15in}
\noindent\large\textbf{2. Vortex Filament Self-Advection with Slipping Fluid Motion}\\

Consider a vortex filament embedded in a fluid which is in a slipping motion with respect to the fluid. The velocity of an element of the vortex filament is then given by

\beq
{\bf v}=\gamma \kappa{\bhat} + U\that
\eq

\noindent where $\that,\nhat$ and $\bhat$ are unit tangent, unit normal and unit binormal vectors, respectively, to the vortex filament; $\kappa$ is the curvature and $\gamma$ is the strength of the vortex filament, and $U$ is the slipping fluid speed. We assume that the slipping motion does not cause changes in the shape of the vortex filament in time (Kida \cite{kid}).

We first note the Frenet-Serret formulae for the differential geometry underlying the vortex filament,

\beq
{\bf x^\pr}=\that,\that^\pr=\kappa\nhat,\nhat^\pr=\tau{\bhat}-\kappa\that,{\bhat}^\pr=-\tau\nhat
\eq

\noindent where $\tau$ is the torsion and primes denote differentiation with respect to the arc length $s$. Equation (2) lead to 

\beq
\lp{\nhat} + i{\bhat}\rp^\pr=-i\tau\lp{\nhat}+ i{\bhat}\rp-\kappa\that
\eq

\noindent which suggests, it is pertinent to introduce, following Hasimoto \cite{has}, 

\beq
\left.
\begin{matrix}
\begin{aligned}
{\bf N}&\equiv\lp{\nhat}+i\bhat\rp e^{i\underset{0}{\overset{s}\int}\tau\lp s,t\rp ds}\\
\\
\psi&\equiv \kappa\lp s,t\rp e^{i\underset{0}{\overset{s}\int}\tau\lp s,t\rp ds}
\end{aligned}
\end{matrix}\rc
\eq

\noindent (4) implies,

\beq
{\bf N}\cdot{\that}=0,{\bf N}\cdot{\bf N}=0,{\bf N}\cdot{\bf\overline{N}}=2
\eq
	
\noindent where the bar overhead denotes the complex conjugate of the quantity in question. (4) further leads to 

\beq
{\bf N}^\pr=-\psi\that
\eq 

\noindent and 

\beq
{\that}^\pr=Re\lp\psi{\bf\overline N}\rp=\fhalf\lp\psi{\bf\overline N}+\overline{\psi}{\bf N}\rp.
\eq

On the other hand, we have from equations (1), (2) and (4), 

$$\dot{\bf\that}=\gamma\lp \kappa^\pr{\bhat}-\kappa\tau\nhat\rp+\frac{U}{2}\lp\psi{\bf\overline N}+\overline{\psi}{\bf N}\rp+U^\pr\that$$

\noindent or

\beq
{\dot\that}=i\frac{\gamma}{2}\lp\psi^\pr{\bf\overline N}-\overline{\psi}^\pr{\bf N}\rp+\frac{U}{2}\lp\psi{\bf\overline N}+\overline{\psi}{\bf N}\rp+U^\pr\that.
\eq
	
\noindent Putting, in accordance with equations (5), 

\beq
\dot{\bf N}=\al {\bf N}+\beta\that
\eq

\noindent we have

\beq
\al+\overline\al=\fhalf\lp\dot{\bf N}\cdot\overline{\bf N}+\dot{\overline{\bf N}}\cdot{\bf N}\rp=\fhalf\frac{\pa}{\pa t}\lp {\bf N}\cdot\overline{\bf N}\rp=0
\eq

\indent from which,

\beq
\al=iR
\eq

\noindent $R$ being a real-valued scalar function.

Next, we have, on using equation (8),

\beq
\beta=\dot{\bf N}\cdot{\that}=-{\bf N}\cdot\dot{\that}=-i\gamma\psi^\pr-U\psi.
\eq

Using (11) and (12), (9) becomes

\beq
\dot{\bf N}=iR{\bf N}-\lp i\gamma\psi^\pr+U\psi\rp\that.
\eq

\vspace{.15in}
Taking the time-derivative of equation (6) and the arc length-derivative of equation (13), and using equations (7) and (8), we obtain

$$\dot{\bf N}^\pr=-\psi{\dot\that}-\dot{\psi}\that$$

\noindent or

\beq
\dot{\bf N}^\pr=-\psi\lb i\frac{\gamma}{2}\lp \psi^\pr\overline{\bf N}-\psi^\pr{\bf N}\rp+\frac{U}{2}\lp\psi\overline{\bf N}+\overline{\psi}{\bf N}\rp+U^\pr\that\rb-\dot{\psi}\that
\eq
	
\noindent and

\beq
\begin{matrix}
\begin{aligned}
\dot{\bf N}^\pr&=iR^\pr{\bf N}-iR\psi{\that}-i\gamma\lb\psi^{\dpr}{\that}+\fhalf\psi^\pr\lp\psi\overline{\bf N}+\overline{\psi}{\bf N}\rp\rb\\
&-\lp\psi U^\pr+U\psi^\pr\rp{\that}-\fhalf U\psi\lp\psi\overline{\bf N}+\overline{\psi}{\bf N}\rp.
\end{aligned}
\end{matrix}
\eq

\noindent Equating (14) and (15), we obtain

\beq
\dot{\psi}=\gamma\psi^{\dpr}+iR\psi+U\psi^\pr
\eq

\beq
i\frac{\gamma}{2}\psi\overline{\psi}^\pr=iR^\pr-i\frac{\gamma}{2}\psi^\pr\overline{\psi}.
\eq

Equation (17) leads to 

\beq
R=\frac{\gamma}{2}\lp |\psi|^2+A\rp
\eq

\noindent $A$ being an arbitrary constant. 

Using (18), equation (16) becomes

\beq
\frac{|}{i}\lp \dot{\psi}-U\psi^\pr\rp=\gamma\psi^{\dpr}+\frac{\gamma}{2}\lp |\psi|^2+A\rp.
\eq

Looking for a solution of the form, 

\beq\tag{20a}
\psi\lp s,t\rp=\kappa\lp\xi\rp e^{i\underset{0}{\overset{s}\int}\tau\lp\xi\rp ds}
\eq

\noindent where,

\beq\tag{20b}
\xi\equiv s-ct.
\eq

\noindent We then obtain

\beq\tag{21}
\lp c+U\rp \kappa^\prime=2\gamma \kappa^\prime\tau+\kappa\gamma\tau^\prime
\eq

\beq\tag{22}
\begin{matrix}
\begin{aligned}
-c \kappa & \lb \tau \lp \xi \rp -\tau \lp -ct \rp\rb -U\kappa \tau=\gamma \lb \kappa^{\dpr} -\kappa \tau^2 + \fhalf \lp \kappa^2+A \rp \kappa \rb.
\end{aligned}
\end{matrix}
\eq

Assuming $U=const$, equation (21) leads to 

\beq\tag{23}
\lb \lp c+U\rp -2\gamma\tau \rb \kappa^2=0
\eq

\noindent from which, 

\beq\tag{24}
\tau =\frac{c+U}{2\gamma}=const
\eq

\noindent which implies that the velocity of propagation of the wave along the filament, in the observer's frame, is twice the torsion.

Using (24), equation (22) leads to 

\beq\tag{25}
\kappa^{\dpr}-\lp\tau^2-\frac{1}{\gamma}U\tau+\frac{A}{2}\rp \kappa+\fhalf \kappa^3=0
\eq

Putting,

\beq\tag{26}
A=\frac{c^2-\nu^2}{2\gamma^2}
\eq

\noindent equation (25) becomes

\beq\tag{27}
\kappa^{\dpr} -\left(\dfrac{\nu^2-U^2}{4\gamma^2}\right) \kappa+\fhalf \kappa^3=0
\eq

\noindent which leads to the modified Hasimoto soliton solution, 

\beq\tag{28}
\kappa\lp\xi\rp=\frac{1}{\gamma}\sqrt{\nu^2-U^2}~ \text{sech}~\lp\frac{1}{2\gamma}\sqrt{\lp\nu^2-U^2\rp}\xi\rp.
\eq

\noindent (28) shows that a strong slipping motion $\lp U^2>\nu^2\rp$ prevents the existence of the Hasimoto soliton.

The case $A=O$, from (26), corresponds to $\nu =c$, and on using (24), (28) then becomes

\beq\tag{29}
\kappa\lp\xi\rp=2\sqrt{\frac{\tau}{\gamma}\lp\gamma\tau-U\rp}~\text{sech}~\lp\sqrt{\frac{\tau}{\gamma}\lp\gamma\tau-U\rp}\xi\rp.
\eq

\noindent (29) shows that the critical slipping speed (above which the Hasimoto soliton fails to exist) is equal to the torsion (as may also be discerned from (24)). This is in total agreement with the existence condition for the helical-twist soliton deduced previously via the reformulation of the Da Rios-Betchov equations in the \tex{extrinsic} vortex filament coordinate space (Shivamoggi and van Heijst \cite{shi}).

\vspace{.20in}

\noindent\large\textbf{3. Discussion}\\
	
It may be mentioned that the Hasimoto soliton (namely, (28), in the limit $U\Rightarrow O$) is in good qualitative agreement with laboratory experiment observations of helical-twist solitary waves propagating on concentrated vortices in rotating fluids (Hopfinger et al. \cite{hop}), even though some of the assumptions underlying LIA (like the lack of coupling of the vortex motion to the degrees of freedom of the vortex core) are violated in this experiment. A vortex filament immersed in a non-ideal fluid, thanks to viscous diffusion, experiences a slipping motion with respect to the fluid. In recognition of this, in this paper, the effect of this slipping motion on the Hasimoto soliton is investigated, and an \tex{exact} solution is given  to describe this process. A strong slipping motion is shown to prevent the existence of the Hasimoto soliton. The critical slipping speed (above which the Hasimoto soliton fails to exist) is shown to be equal to the torsion, in agreement with the the existence condition for the helical-twist soliton deduced previously via the reformulation of the Da Rios-Betchov equations in the extrinsic vortex filament coordinate space (Shivamoggi and van Heijst \cite{shi}).

\vspace{.15in}

\noindent\large\textbf{Acknowledgements}\\

Most of the work reported in this paper was done at the Eindhoven University of Technology during my visiting appointment supported by The Netherlands Organization for Scientific Research (NWO). I am very thankful to Professor Gert Jan van Heijst for his immense hospitality and helpful discussions.

\end{document}